\documentclass[review,12pt]{elsarticle}

\usepackage{tabto}
\usepackage{amssymb}
\usepackage{pdflscape}
\usepackage{color}
\usepackage{multirow}
\usepackage{graphicx}
\usepackage{subcaption}
\usepackage{url}
\usepackage[table]{xcolor}
\usepackage[justification=centering]{caption}
\usepackage{geometry}

\usepackage{listings}

\begin{document}

\begin{frontmatter}



\title{\textbf{\textit{gh0stEdit}}: Exploiting Layer-Based Access Vulnerability Within Docker Container Images}


\author{Alan Mills}
\author{Jonathan White}
\author{Phil Legg}
\affiliation{organization={Computer Science Research Centre},
            addressline={University of the West of England}, 
            city={Bristol},
            country={UK}}

\begin{abstract}
Containerisation is a popular deployment process for application-level virtualisation using a layer-based approach. Docker is a leading provider of containerisation, and through the Docker Hub, users can supply Docker images for sharing and re-purposing popular software application containers. Using a combination of in-built inspection commands, publicly displayed image layer content, and static image scanning, Docker images are designed to ensure end users can clearly assess the content of the image before running them. In this paper we present \textbf{\textit{gh0stEdit}}, an exploit that fundamentally undermines the integrity of Docker images and subverts the assumed trust and transparency they utilise. The use of gh0stEdit allows an attacker to maliciously edit Docker images, in a way that is not shown within the image history, hierarchy or commands. This attack can also be carried out against signed images (Docker Content Trust) without invalidating the image signature. We present a detailed case study for this exploit, and showcase how gh0stEdit is able to poison an image in a way that is not picked up through static or dynamic scanning tools. We highlight the issues in the current approach to Docker image security and trust, and expose an attack method which could potentially be exploited in the wild without being detected. To the best of our knowledge we are the first to provide detailed discussion on the exploit of this vulnerability.
\end{abstract}


\end{frontmatter}

\section{Introduction}

%
%
%
%

Docker provides a lightweight solution for software distribution that packages up all software dependencies in a single container image so that it can be readily deployed in other computing environments. As a technology paradigm, since its introduction in 2013, it has attracted widespread adoption in both cloud computing systems and local deployments, and is regarded as an industry standard.

Docker Hub provides a readily available repository of many common software environments. Many developers utilise these images as the basis for further application development. A key component of the Docker environment is the use of image layers, whereby layers of functionality are built up in a modular fashion. For example, a container image may have a Linux distribution as the base image, which then has a Python environment built upon this, followed by software library dependencies, and then followed by the application source code.

In this paper, we detail the exploitation of a vulnerability within the Docker framework that we refer to as \textbf{\textit{gh0stEdit}}. The vulnerability essentially allows the manipulation of the image layers within a Docker container to embed malicious content in a manner that would be indistinguishable from a benign image when using common industry-standard reporting and scanning techniques. We also test this attack on an image signed using the Docker Content Trust (DCT) and show how the image signature is not invalidated. We believe that this is the first report detailing the exploitation of this vulnerability, in which the integrity of an image can be severely compromised.

We outline the vulnerability and develop a proof-of-concept to illustrate how this can be exploited as an attack vector. We conduct further analysis using a variety of tools designed for inspecting Docker containers, and show that existing toolchains fail to recognise the modification and show little to no change from the benign container images. 

We believe that this poses a critical security risk for developers and Continuous Integration and Continuous Deployment (CI/CD) pipelines using available container images. Docker Hub has reported over one billion downloads for some of the most popular container images that would be used for further development. We have disclosed our findings to Docker and have logged a Common Vulnerabilities and Exposures (CVE) report with MITRE, to ensure that this vulnerability can be resolved and mitigated against in the future.

\section{Background}
\label{subsec:technical_background}

Containers are a lightweight form of virtualisation, with a focus on the application layer. A container will include an application and all the required libraries and packages to run that application. Additional packages and utilities are often omitted to reduce ``bloat'', and it relies on the host OS for OS-level components, such as the kernel.

\begin{figure}[t]
\centering
\includegraphics[width=8cm, trim={100 0 100 0}, clip]{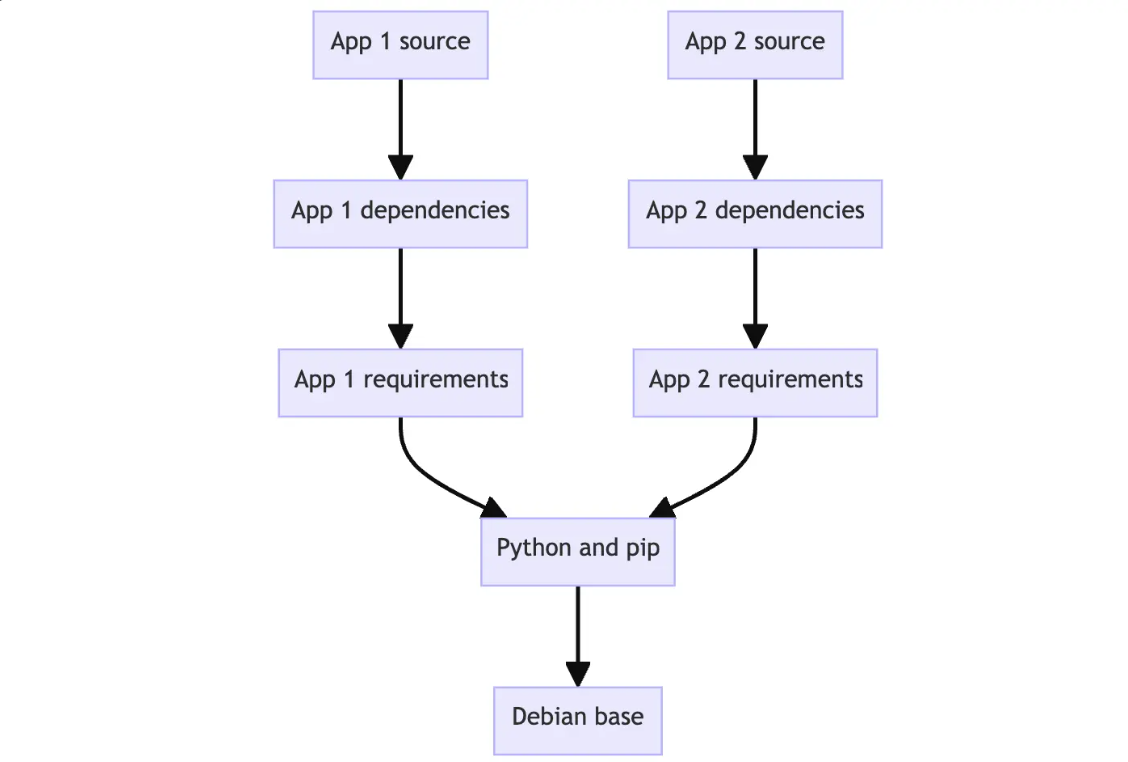}
    \caption{Example of two containers with shared layers}
    \label{fig:docker_shared_layers}
\end{figure}

Containers are packaged as images, which are read-only templates that become containers at runtime. These images are built utilising layers. Multiple containers or images are able to share common base images and layers, which therefore reduces duplication of commonly used layers\footnote{https://docs.docker.com/get-started/docker-concepts/building-images/understanding-image-layers/} (as illustrated in Figure~\ref{fig:docker_shared_layers}).

\begin{figure*}[t]
\centering
\includegraphics[width=\textwidth]{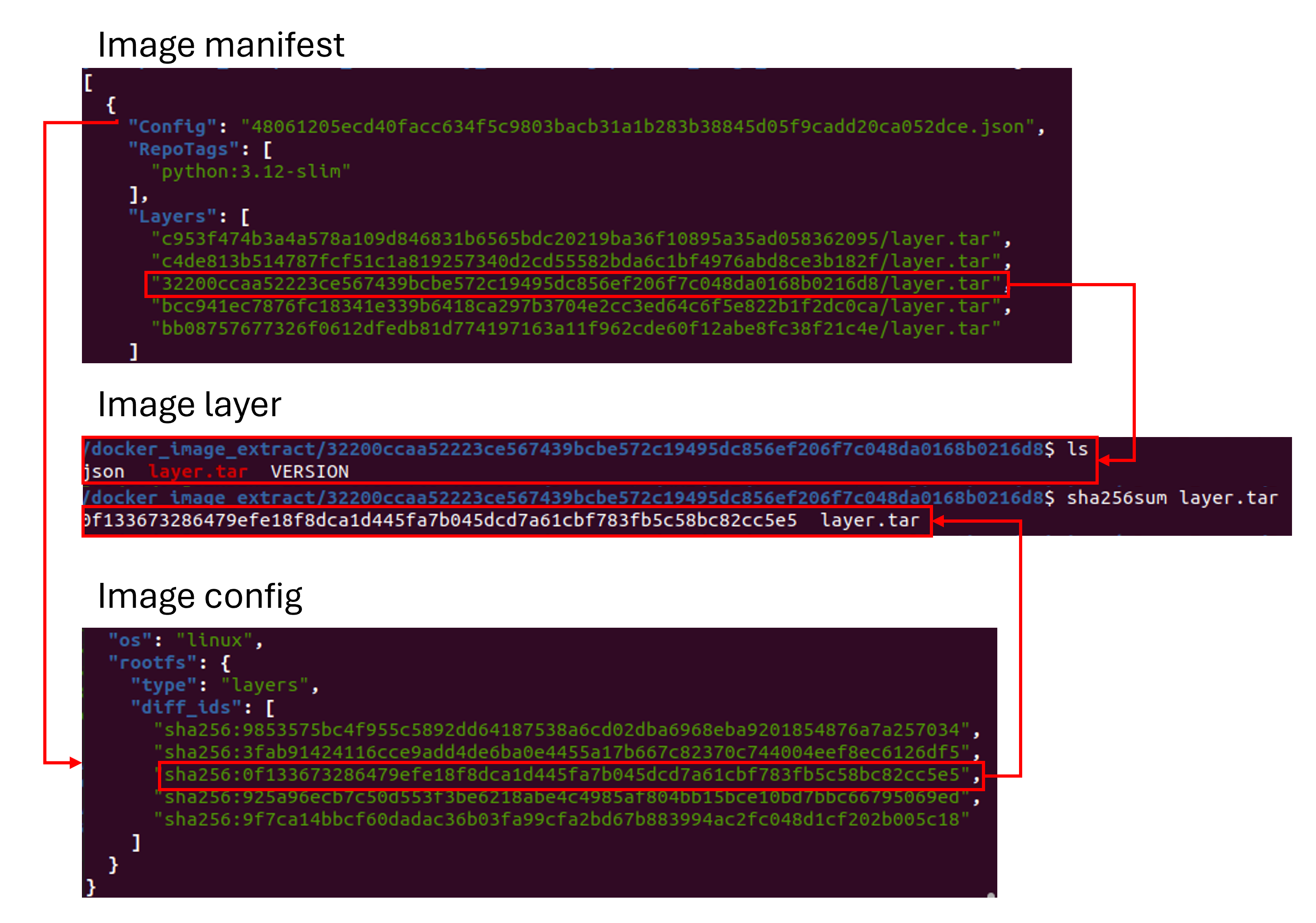}
    \caption{Breakdown of a Docker image manifest, config and layer relationship}
    \label{fig:manifest_config_layer}
\end{figure*}

This, along with the focus on keeping installed packages to the required minimum, ensures that containerisation provides a lightweight virtualisation method. While images are commonly `pulled' from online repositories, they can also be saved as archive files\footnote{\url{https://docs.docker.com/reference/cli/docker/image/save/}} which can be distributed and subsequently loaded onto other hosts. In either format the images themselves are composed of layers, with each image also containing a manifest.json that lists the layers that make up the image. In Docker images there is also a linked config JSON file which is listed in the manifest. This config will contain details for the image runtime, such as commands, environment variables, and the construction of the root filesystem. This is provided as an array of layers, with each layer represented by the sha256sum of the associated layer.tar. This sha256sum represents the checksum of the content of this layer, using SHA256 and is part of the layer verification process to ensure filesystem integrity~\footnote{\url{https://github.com/moby/moby/blob/master/pkg/tarsum/tarsum_spec.md}}. If there is a mis-match between the layers SHA256 checksum and the associated layer entry in the array then the image will fail verification. Figure~\ref{fig:manifest_config_layer} shows the relationship between the manifest, config, layers and associated use of sha256sum, which have been taken from an extracted image archive file. 

The \textit{layer.tar} contains all the files and changes within that layer. In the highlighted example we can see that the SHA256 checksum for the  \textit{layer.tar} is the associated entry for the layer in the image config. The JSON also details the parent layer, in the example shown in Figure~\ref{fig:manifest_config_layer}, the parent for the highlighted layer would be: 
\newline {\footnotesize c4de813b514787fcf51c1a819257340d2cd55582bda6c1bf4976abd8ce3b182f}.

Since the layers are an ordered array, the last (most recent) layer in any image is always {\tt layers[-1]}. We can therefore see that the last layer within this image would be: {\footnotesize bb08757677326f0612dfedb81d774197163a11f962cde60f12abe8fc38f21c4e}.

The ordering of layers is important, as a change in one layer can be overwritten by subsequent layers. For example, the installation of a package at layer 3 may then be overwritten by its removal or replacement in layer 4. When taken into context for the gh0stEdit, this meant that initial Proof of Concept (PoC) code was created to automatically effect changes at the last layer ({\tt layers[-1]}) to avoid any overwrites. However, it is possible to make edits within earlier layers which can ensure that the changes are less obvious and blend in with similar changes occurring at the target layer. For example, if we are adding or replacing a binary within the {\tt /usr/local/bin} folder, doing so at the latest layer will make this change more apparent during deep layer inspection using tools like Dive~\cite{gitdive}, especially if no other changes at this location occur in this layer. However, by making this change at an earlier layer, but ensuring that this was the last layer to make changes to this folder, our own edit blends into the existing changes and will correlate with the associated layer commands. An example of this is shown in Section~\ref{sec:vuln}.

\section{Related Works}

\label{sec:related_works}

The effectiveness of container vulnerability scanning has often been discussed with a focus on the number and severity of identified CVEs. Though there has been discussion on the accuracy of different scanning tools, this is in the context of missed CVEs or differing severities within images, rather than the overall use case and functionality of vulnerability scanning. The combination of trusted images and regular scanning is often presented as a key security measure for mitigating image vulnerabilities and poisoned images.  

In this section we examine existing works within container cyber security, with a focus on container security and scanning solutions, as well as current research on supply chain and container attacks. We discuss the currently recommended safeguards, which we then test against in our case studies, and highlight how our attack fits within previously discussed but unexplored attack scenarios. 

\subsection{Container Security and Scanning software (OS)}

Container security has been a prominent issue for many years. Thanh Bui examined the challenges associated with Docker security in their paper Analysis of Docker Security in 2015~\cite{bui2015analysis}. This study focused on the internal security mechanisms of Docker, as well as its interaction with the Linux kernel. Bui analysed Docker's isolation capabilities, identifying concerns related to shared network resources, direct kernel access and the potential impact of running ``privileged" containers. 

Since that time, there has been a marked shift in focus towards the security implications and vulnerabilities of container images themselves, reflecting the evolving nature and changing concerns within container based security.

In the paper, \textit{Mitigating Docker Security Issues}, Robail Yasrab looked to ``outline some significant security vulnerabilities at Docker and counter solutions to neutralise such attacks''~\cite{yasrab2018mitigating}. Yasrab explored defences against various attacks, including poisoned images, and proposed strategies for mitigating these risks. These strategies included the use of trusted and signed images through Docker Content Trust, alongside regular security audits. Similarly, in the paper, \textit{Container security: Issues, challenges, and the road ahead}, Sultan \textit{et al.} emphasised the necessity of ``periodic vulnerability scanning" for container images and recommended the use of trusted images only, ``verifying the images using signatures''~\cite{sultan2019container}. The authors also stressed the importance of dynamic and runtime scanning for container applications.

Liu \textit{et al.} studied CVEs within Docker images, in their paper entitled \textit{Understanding the Security Risks of Docker Hub}, as well as the prevalence of malicious images and sensitive parameters \cite{liu2020understanding}. Their proposed framework employs a combination of Anchore and the VirusTotal API to identify CVEs and malicious files, respectively. The presence of a ``malicious executable" is used as criterion to confirm that an image is malicious, with an emphasis on the analysis of ``executed programs", which are identified through the images entry file. The authors acknowledge the limitation of their findings, noting that VirusTotal may fail to detect certain malware.

In their survey paper, \textit{Threat Modelling and Security Analysis of Containers: A Survey}~\cite{wong2021threat}, Wong \textit{et al.} examine unresolved security issues within container environments. Using the STRIDE framework, they identified 12 vulnerabilities related to container security, including issues surrounding image tampering and its subsequent impact on CI/CD pipelines. The authors also discuss existing mitigation strategies for these vulnerabilities, such as the use of scanning tools like Docker Scout and Anchore (now superseded by Grype) to perform code scans at each stage of the build process. While they acknowledge the limitations of scanning tools, their focus is primarily on the number of vulnerabilities that remain undetected, rather than the limitations of current scanning methodologies and the possibility that changes to an image may be entirely overlooked by scans.   

Another approach to ensuring container image security is explored by the authors of Confine: Automated system call policy generation for container attack surface reduction. Their research focuses on syscall monitoring within micro-services~\cite{ghavamnia2020confine}, particularly addressing kernel vulnerabilities and preventing runtime escapes. However, their solution requires user input and is designed to capture syscalls during an initial dynamic analysis and monitoring phase. In cases where an image has been compromised by poisoning prior to deployment, malicious syscalls will be included within those initially captured. This is particularly concerning where the main executable has been compromised, and the attack vector has been carefully crafted to align with the applications expected functionality.

The security measures and safeguards discussed in the aforementioned papers are all explored within our attack case studies, where we demonstrate how gh0stEdit is capable of evading both static and dynamic scans. Furthermore, we have successfully altered a signed image without invalidating its signature, resulting in a poisoned image that would pass through these security safeguards without detection. 

\subsection{Supply Chain and Container Attacks}

In their paper, Ladisa \textit{et al.} created a taxonomy of attacks on Open Source Software (OSS) \cite{ladisa2023sok}. While their primary focus was on OSS packages, they also examined the impact of supply chain attacks on container images and Docker Hub. Their study considered both attacks and safeguards, including responses from developers and maintainers, referred to in the paper as ``domain experts''. Among the safeguards discussed were the use and maintenance of a Software Bill Of Materials (SBOM), generated through automated tools, the careful inspection of changes during the build process and integration of scanner tools within the CI/CD pipeline. 

In a related paper, \textit{Exploring the Threat of Software Supply Chain Attacks on Containerised Applications}~\cite{mounesan2023exploring}, the authors investigate ``container susceptibility to security issues intentionally introduced by malicious actors''. They focus on the impact of package vulnerabilities within the container ecosystem and the potential to serve as a vector for supply chain attacks. They use the SBOM generated by Docker for each container. However, they note that ``certain dependencies installed through commands in Docker files, as well as application dependencies, may not be listed in the SBOM''. Although this gap can be mitigated by inspecting Docker files, in our attack scenario, neither the Docker files nor the the generated SBOM reveal the presence of altered binaries or indications of compromise. Our attack examples also successfully evaded detection by both static and dynamic vulnerability scanners, allowing this form of supply chain attack to bypass the suggested safeguards.

Tomaer \textit{et al.} focus on Docker based attacks in their work, \textit{Docker security: A threat model, attack taxonomy and real-time attack scenario of DoS}~\cite{tomar2020docker}. They emphasise that security is a ``crucial concern" in the use of containerisation and outline several attack scenarios. Within their attack taxonomy, they look at the threat posed by poisoned images, highlighting the vulnerability of signed manifests due to the lack of authentication from Docker. They note that ``an attacker with a signed manifest can transmit any image which can lead to serious vulnerabilities''~\cite{tomar2020docker}. However, their work includes only a limited case study, focusing primarily on a denial-of-service (DoS) attack within the Docker environment. It should also be noted that their work is a later, academic publication, that covers issues previously discussed by Jonathan Rudenberg~\cite{rudenberg2015} who highlights weaknesses within the Docker image verification process in 2015. This includes issues around the use of tarsum, the ability to unpack data from the container image and the lack of properly validated image manifest checking and verification.  

Our work extends the findings of these previous studies by demonstrating how a Docker image can be poisoned in a manner that evades currently recommended safeguards. We also provide a detailed case study that builds upon previously identified attack scenarios. To the best of our knowledge, this is the first paper to identify and exploit this specific attack methodology.

\section{\textit{gh0stEdit} Exploit}
\label{sec:vuln}

\begin{figure}[!th]
\centering
\includegraphics[width=\textwidth]{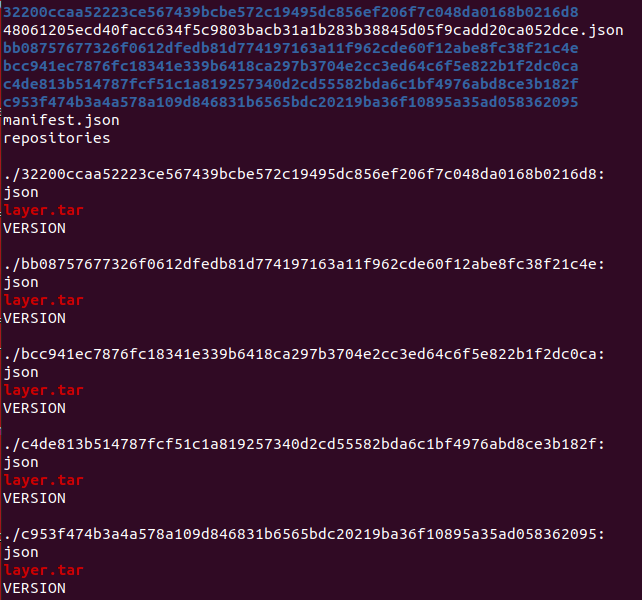}
    \caption{Extracted archive content of Python 3.12-slim container image (EXT4 filesystem)}
    \label{fig:file-archive}
\end{figure}

The exploit discussed in this paper pertains to the ability to access and modify the raw layers of the container image. Once an image from Docker Hub has been pulled, it can be saved and accessed as a {\tt tar} archive through the terminal: {\tt docker save -o python.tar python:3-12-slim}. The extracted archive then provides access to all individual layers. Figure \ref{fig:file-archive} illustrates the content of an extracted image archive in a Linux system. 

On Linux-based systems, the extracted image archive includes a manifest.json, which contains hashes specifying the layers and their order for deployment. These layers may consist of bash commands or binary content, to construct the container.

\begin{figure}[!t]
\centering
\includegraphics[width=\textwidth]{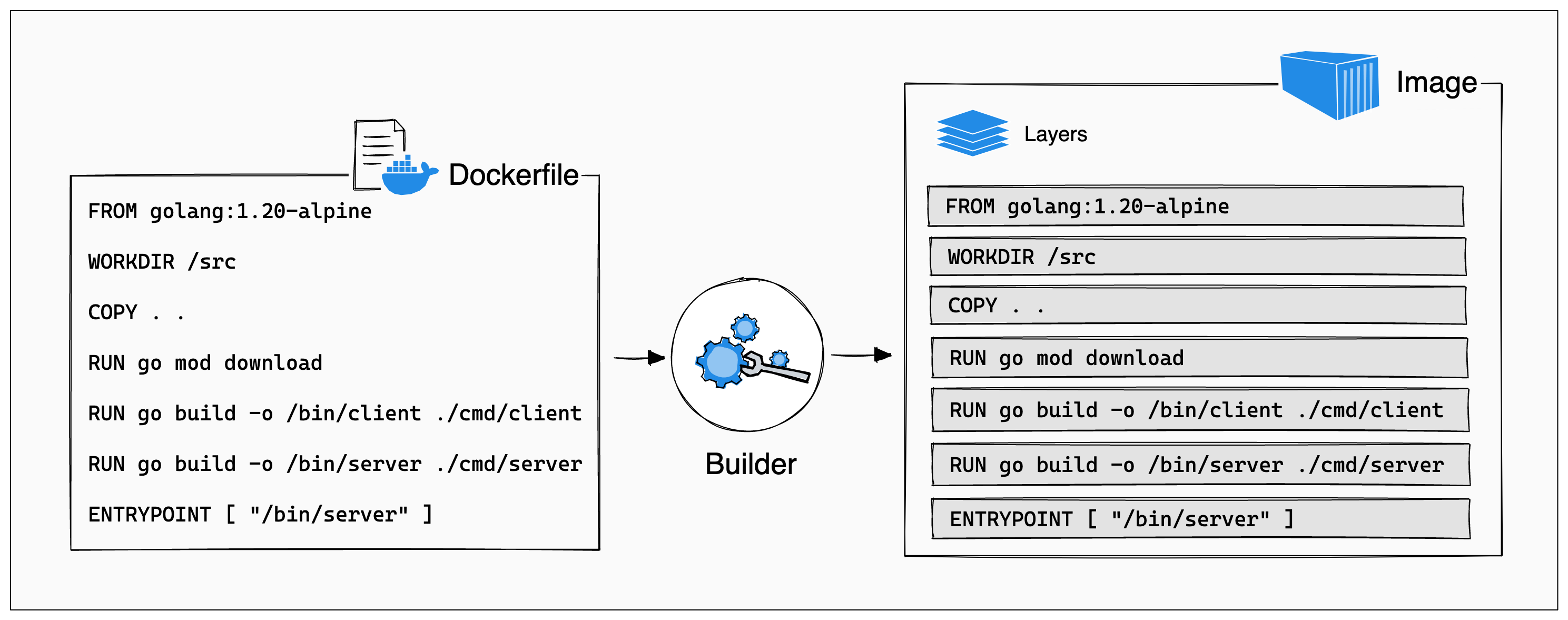}
    \caption{Example of Docker layers to illustrate copying new file content into a layer}
    \label{fig:docker_layers}
\end{figure}

Figure~\ref{fig:docker_layers} illustrates the concept of Docker layers\footnote{https://docs.docker.com/build/guide/layers/}. Specifically, in this example the Dockerfile instructions copy content into the container image. Consequently, the corresponding layer for this instruction will contain the new files and folders added by the COPY instruction.

\begin{figure*}[t!]
  \centering 
  \subfloat[gh0stEdit image]{\includegraphics[width=0.45\textwidth]{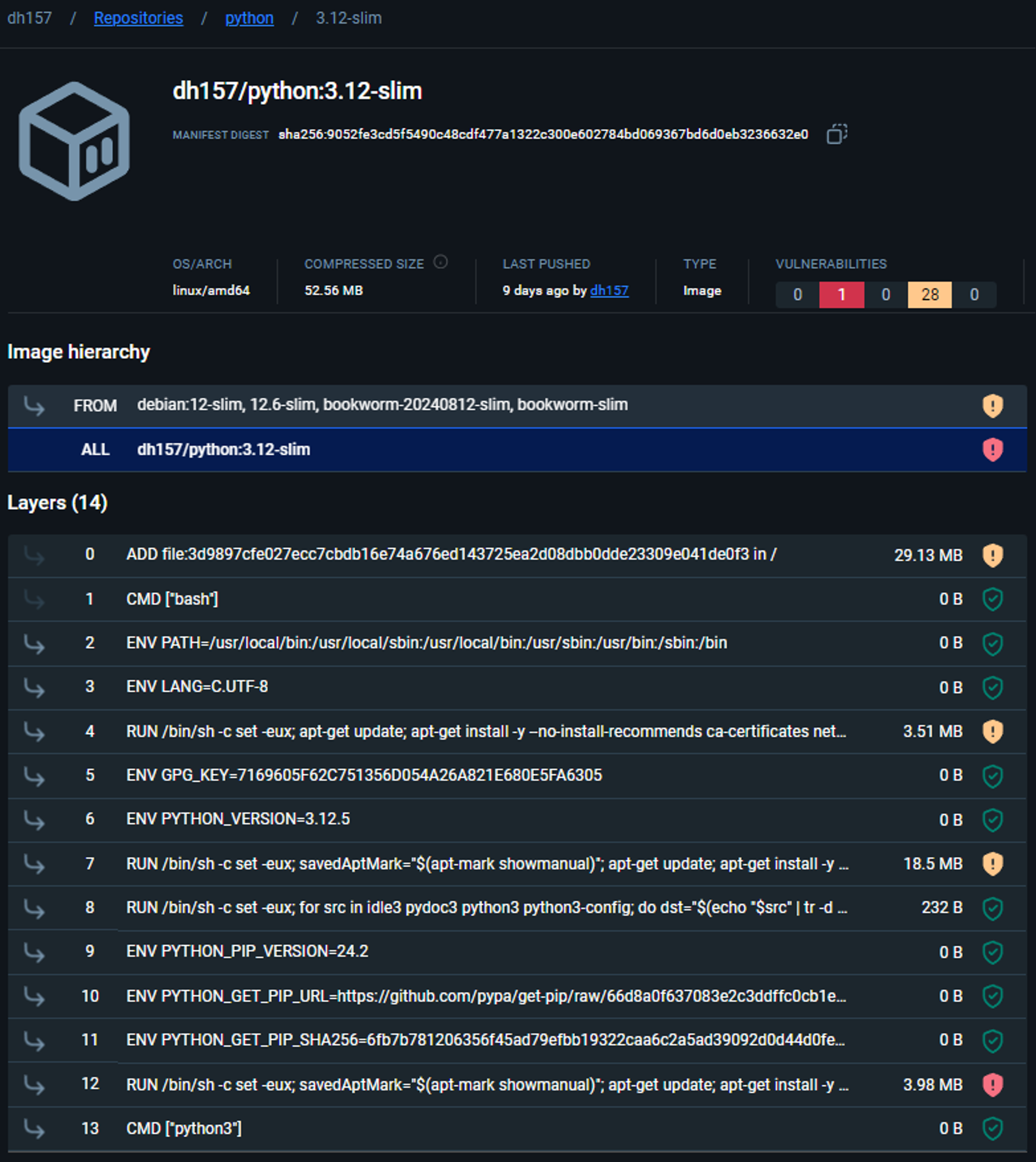}} 
  \qquad 
  \subfloat[original Python 3.12-slim]{\includegraphics[width=0.45\textwidth]{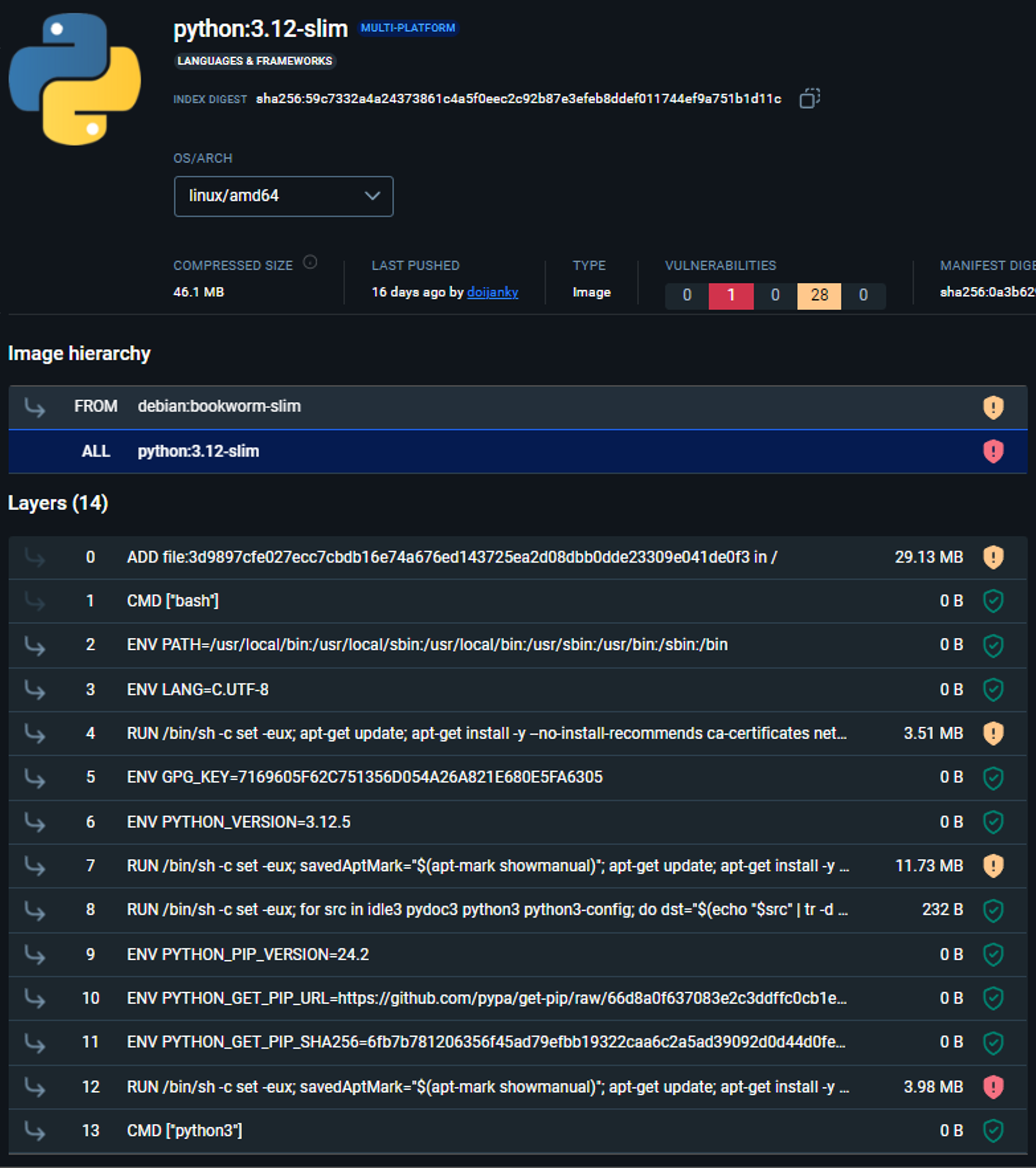}} 
  \caption{Docker image hierarchy (a) gh0stEdit image and (b) original Python 3.12-slim (as of 29th August 2024).}
  \label{fig:layer_comparision}
\end{figure*}

This layer-based approach offers significant utility, as discussed in Section~\ref{subsec:technical_background}, as well as transparency. Users can observe modifications made at each layer using built-in Docker functionality (e.g., {\tt docker history}), third party tools (e.g., Dive~\cite{gitdive}), and the image hierarchy view as displayed on the Docker Hub webpage~\cite{dockerhub} (Figure~\ref{fig:layer_comparision}). Any changes to files, whether they are added, removed or modified, results in the creations of a new layer, which is recorded as part of the image history. Additionally, Docker offers a squash functionality~\cite{dockersquash}, which can merge all layers into a single one. While this could be used to obfuscate changes within an image, the use of such functionality may appear suspicious, prompting further investigation, or causing an image to be deemed untrustworthy by the community. 


In contrast, our approach edits the content of the layer directly within the extracted image archive. This method prevents changes to the container being reflected in the image hierarchy. Changes can be targeted at specific layers so that alterations blend with other functionality present at the same level. Consequently, identifying such modifications using layer-based inspection tools such as Dive become more difficult. For example, if specifically targeting the folder ({\tt /usr/bin/local}) then there will be multiple modifications present in the original image. In the original image the python3 is merely a symbolic link, whereas in our edited compromised image,  it appears as a binary executable. This subtle alteration would likely be unnoticed and is easy to overlook. Especially without a means of direct comparison. The nature of this attack vector is explored further in Section~\ref{sec:attack_example_python}. 

\begin{figure}[!t]
\centering
\includegraphics[width=\textwidth]{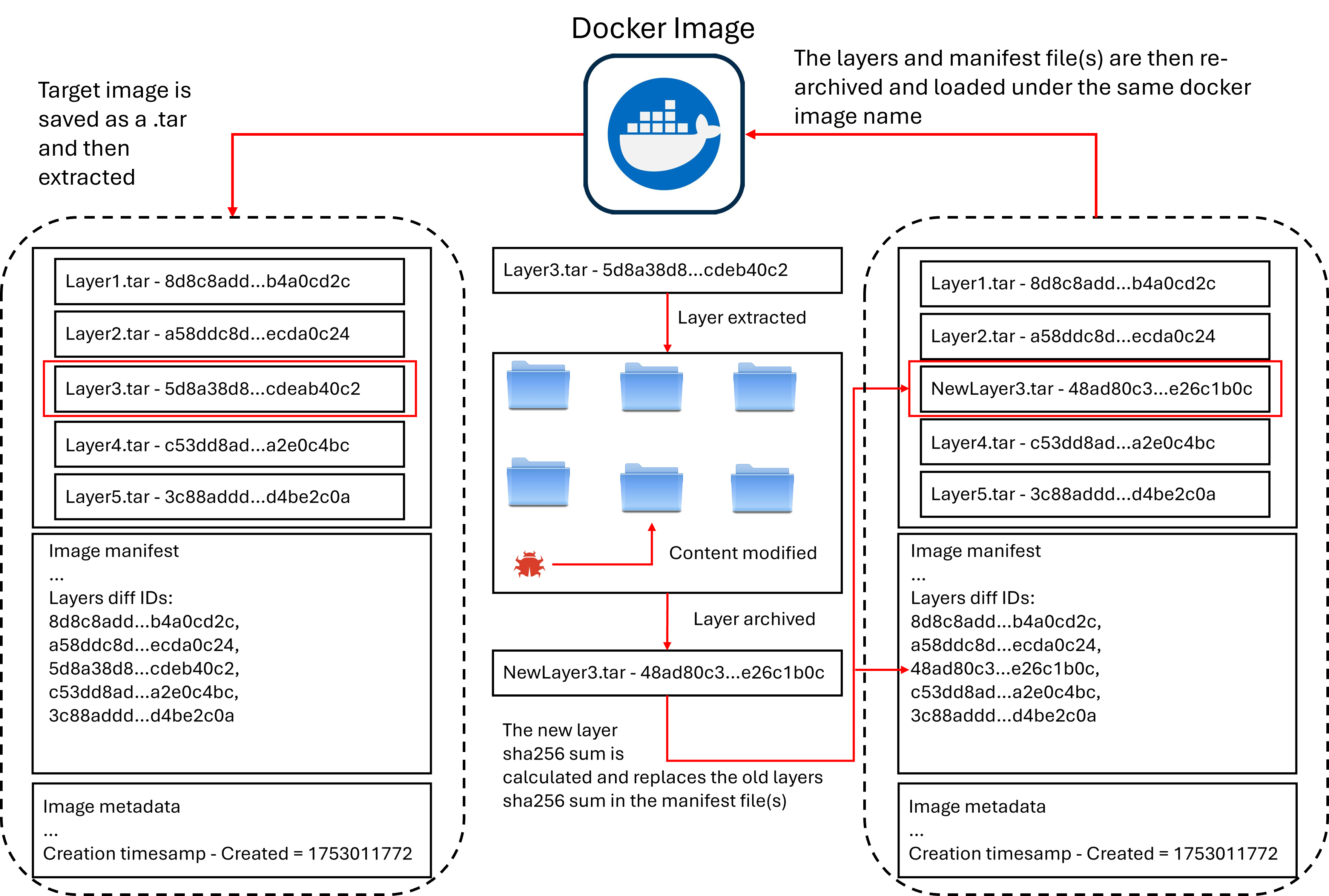}
    \caption{gh0stEdit exploit - Illustrated}
    \label{fig:gh0stEdit_exploit}
\end{figure}

Once the content of a layer has been changed, the attacker updates the sha256sum within the diff\_ids entry of the image JSON configuration. This value is then used to verify the layers being loaded into the Docker image (see Section~\ref{subsec:technical_background}). The image content is then re-archived, complete with the edited layer, and loaded back into the Docker environment, this process is shown in Figure~\ref{fig:gh0stEdit_exploit}. shows Docker will then recognise the change in the layer and overwrite it. Crucially, this change is \textbf{not} visible in the image history or hierarchy. Currently, Docker image metadata does not record time of edits, meaning the image's creation timestamp will remain unchanged. The only indication that a modification has occurred is the altered hash value and minor difference in the file size at the target layer (layer 7), as shown in Figure~\ref{fig:layer_comparision}.

A key issue here is how the layers and image history are handled within Docker images. As part of the image build process the {\tt CreatedBy} value, which contains the commands used to create an image layer~\footnote{https://github.com/moby/moby/blob/master/api/swagger.yaml} are populated during image creation and thereafter read based on existing image metadata. This means that if an image is edited outside the expected Docker build path the image metadata is not updated. This can be achieved through a fairly simple process because, as highlighted by Jonathan Rudenberg~\cite{rudenberg2015}, the image manifest relies on the use of tarsums and allows end users to unpack the image content. This reliance on tarsums means that as long as the check sum for the layers archive file (layer.tar) matches what is in the image manifest, it will be accepted as a legitimate part of the image. It is therefore possible to unpack the image, carry out an edit and then update the associated tarsum, leading to an alteration which passes the image verification process, but bypasses the use of the build process. As such commands like Docker history or inspect will be displaying results based on metadata that was true at the point of image creation.

We demonstrate the impact of this exploit by introducing a malicious binary into the Python environment and downgrading libraries within an existing package, thereby introducing known CVEs that can be subsequently exploited. We describe the attack scenario in detail, along with our analysis of the attacks in the following section. We further build on this attack cases by creating a PoC script which automate the attack, using gh0stEdit to include a reverse shell within images and have it dynamically executed as part of the intended image {\tt Entrypoint}. To conduct our analysis of the attack use cases, we utilise a range of widely-adopted container scanning tools, which are actively used by the community. These tools are summarised in Table~\ref{tab:scanning_tools}. Tools such as Clair, Trivy, Docker Scout and Grype have been selected as they, or the successor tool (Grype being the successor to Anchore), have been used in existing academic research around container security, ~\cite{liu2020understanding},  ~\cite{javed2012understanding}, ~\cite{shu2017study}, ~\cite{wist2021analysis}, ~\cite{chen2022seaf} and ~\cite{mills2022ogma} or to provide functionality that is not covered by these tools to ensure scientific rigour as part of our analysis. For example the use of dynamic analysis provided by NeuVector, non-local image inspection provided by Skopeo or Malware specific detection through the industry standard use of YARA rules or ClamAV scanner. We compare the original base image with the maliciously edited image using each scanner to evaluate whether there are any observable indicators of compromise are present.

Using traditional methods to change a Docker image the addition of new binary files or vulnerable packages will be flagged in the image history and will be highlighted by scanning tools. Using gh0stEdit the changes to an image are not shown, allowing any changes to bypass existing security best practices, something which would be of significant impact in an automated CI/CD process. We expand on these points further in Section~\ref{sec:discussion}.

\begin{table}[!t]
    \scriptsize
    \centering
    \begin{tabular}{ | c | c | c | } 
    \hline
       \textbf{Scanner} & \textbf{Description} & \textbf{Type} \\ \hline
       ClamAV & Malware Scanner & Static  \\ \hline
       Docker Scout & Vulnerability Scanner & Static \\ \hline
       Grype & Vulnerability Scanner & Static \\ \hline
       Skopeo & Image Inspection & Static \\ \hline
       Trivy & Vulnerability Scanner & Static \\ \hline
       YaraHunter & Malware Scanner & Static \\ \hline
       NeuVector & Full Lifecycle Container Security & Dynamic \\ \hline
    \end{tabular}
    \caption{Scanning tools utilised}
	\label{tab:scanning_tools}
\end{table}

\section{Attack Use Case - Malicious Python}
\label{sec:attack_example_python}

In this section we cover an attack use case in detail to highlight steps taken for scanning and detection. We demonstrate how a commonly used binary can be compromised. Using the python:3.12-slim image from the official Python Docker Hub repository as our base, we modify the python3 symbolic link to instead use the compromised binary. The compromised binary functions as a wrapper for the underlying Python installation. However, any argument passed to the wrapper is first sent as the data portion of a web request to a Canarytoken~\cite{canarytokens} before being forwarded to the legitimate python3.12 installation. The modified binary was created using PyInstaller~\cite{pyinstaller}. This approach preserves the intended functionality of the original container image, while introducing a ``poisoned" element in a demonstrable manner. 

\subsection{ClamAV}

\begin{figure}[!t]
\centering
\includegraphics[width=\textwidth]{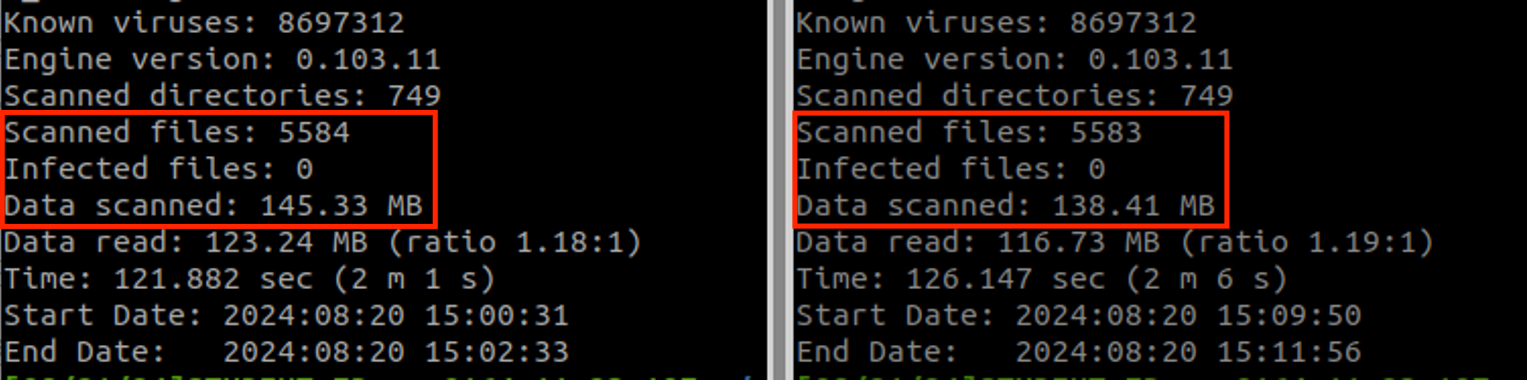}
    \caption{ClamAV scan results - Python image. Left - Scan results for the edited image. Right - Scan results for the base image.}
    \label{fig:clamav_results}
\end{figure}

ClamAV~\cite{clamv} is an Open Source (OS) malware scanner which has served as the foundation for previous container specific malware scanners such as Dagda~\cite{gitdagda}. To perform the scan, both the original base image and the compromised image were exported using Docker export, and extracted. ClamAV was run recursively against these exported filesystems. Crucially, the poisoned binary was not detected, with both scans reporting 0 infected files. The only indication of a difference between the two images was an increase in the number of scanned files and the total amount of data scanned within the edited image, as shown in Figure~\ref{fig:clamav_results}. This discrepancy arises because in the base image python3 is a symbolic link, while in the edited image it is a binary file.  

\subsection{Docker Scout}

\begin{figure}[!t]
\centering
\includegraphics[width=\textwidth]{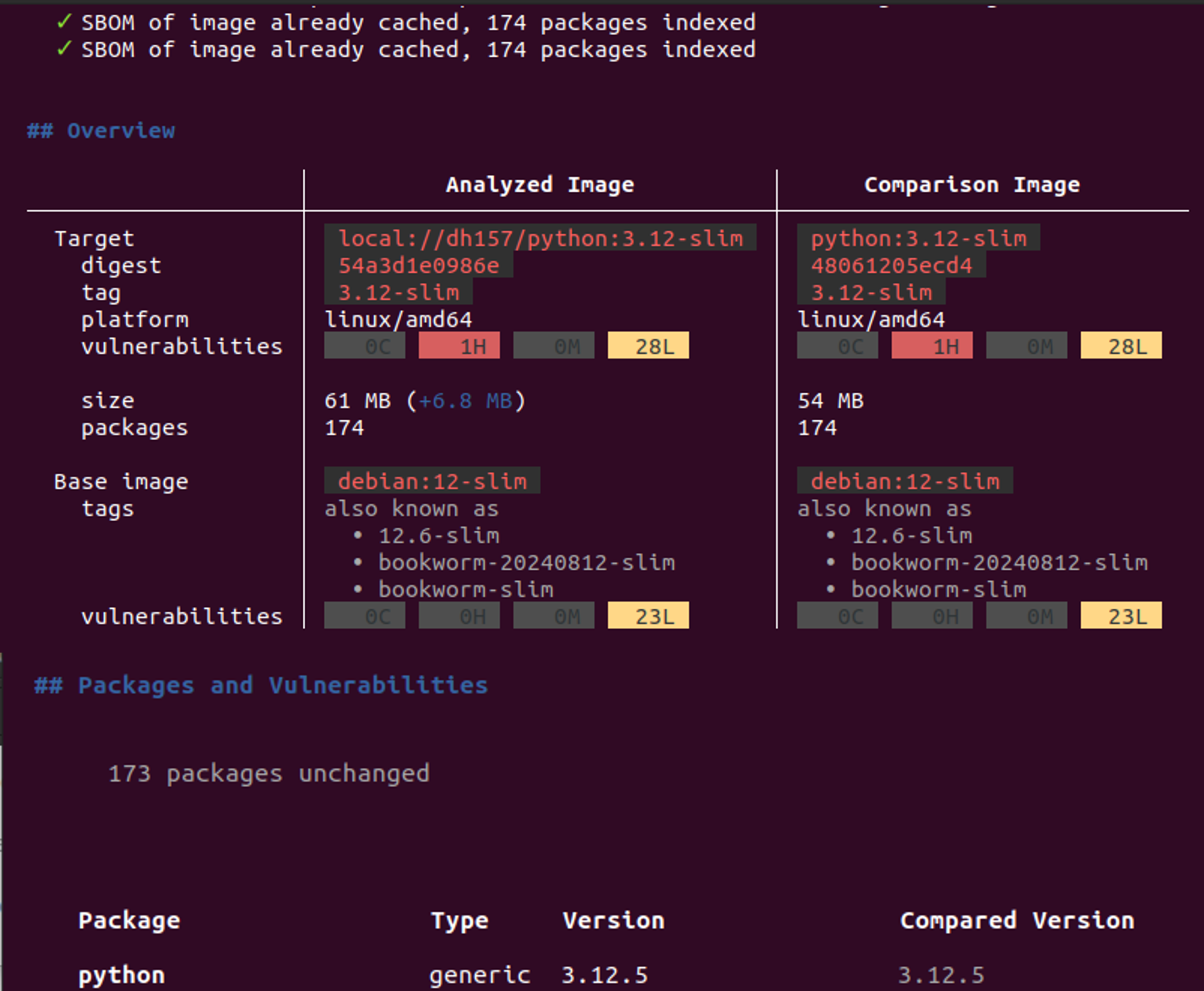}
    \caption{Docker Scout comparative results for the analysed image (compromised) and the comparison image (original). }
    \label{fig:dockerscout_results}
\end{figure}

Docker Scout~\cite{gitdockerscout} is a vulnerability scanner provided by Docker (which replaced Docker Scan~\cite{gitdockerscan}) and is designed to scan images for vulnerabilities and provide comparisons between different images. It can be run locally as a Command Line Interface (CLI) plugin, and can also be enabled within repositories to provide vulnerability information directly within Docker Hub (as shown in Figure~\ref{fig:layer_comparision}). Docker Scout was used as a CLI plugin to compare the edited image with the original base image. As shown in Figure \ref{fig:dockerscout_results}, the scan reported no difference in the packages contained within the two images, with both images displaying the same number of vulnerabilities reported, all of which were related to CVEs present within the base image. Unlike ClamAV, Docker Scout did not report any differences in the number of files or packages between the two images, although as with ClamAV, it did identify a difference in image size.

\subsection{Grype}

Similar to Docker Scout, Grype~\cite{gitgrype} is a vulnerability scanner provided by Anchore. Both the edited and original image were scanned, with no differences reported in terms of vulnerabilities or packages.  Grype's output also includes a base64-encoded manifest and configuration file values. When comparing the scan outputs for both images, differences were observed in these encoded values, along with different hash ID values for the image digest and altered layer. These discrepancies are due to the changed sha256sum values for the edited layer. A difference in image size was also reported.

\subsection{Skopeo}

Skopeo~\cite{gitskopeo} can be used to inspect Docker images, providing similar functionality to the Docker history or inspect commands. It allows users to inspect both local and remote container images prior to downloading. While Skopeo identified differences in size and hash ID for the altered layer when comparing the two images, the commands and image history shown by the tool were identical between them. 

\subsection{Trivy}

Trivy~\cite{gittrivy}, a vulnerability scanner provided by Aqua Security, functions similarly to Docker Scout and Grype. It reported no differences in vulnerability or package between the original and edited images. As with Grype, the difference in hash IDs for the layers and image digests were reported. However, unlike Grype, Trivy does not include manifest or image sizes its output reports. 

\subsection{YaraHunter}

YaraHunter~\cite{gityarahunter} by Deepfence, scans container images for indications of malware using a YARA ruleset and provides details on any matched signatures. In the original image, YaraHunter identified four matches, all associated with SpyEye malware, which were linked to various packages such as pip wheel files and dpkg library files. However, we believe this detection of SpyEye in the base image is highly likely to be a false positive, as there is no indication that the original image includes any malicious content. In the modified image, YaraHunter flagged the altered python3 binary. However, the severity is marked as low, and the binary was flagged due to being created by PyInstaller. An example of the output is shown in Code Listing~\ref{lst:yara_output}. 

\begin{lstlisting}[caption={YaraHunter finding - Edited Python3 binary}, label={lst:yara_output}]
"Matched Rule Name": "MachO_File_pyinstaller",
      "FileSeverity": "low",
      "Full File Name": "/usr/local/bin/python3",
      "rule meta": [
        "",
        "",
        "author : KatsuragiCSL (https://katsuragicsl.github.io) \n",
        "description : Detect Mach-O file produced by pyinstaller \n"
      ],
\end{lstlisting}

We believe it would be highly likely that the association between Python and PyInstaller would mean that many analysts would interpret this detection as a false positive, given that the typical use of PyInstaller for packaging Python binaries. Although it is possible to obfuscate the binary and avoid detection by YaraHunter through a trial and error, we have included this initial match to demonstrate that, even without attempts at obfuscation, the detection could easily be misinterpreted as a false positive. This illustrates that without further context or analysis, detection tools may misidentify legitimate binaries as malicious due to the packaging methods used.

\subsection{NeuVector}

\begin{figure}[!t]
\centering
\includegraphics[width=\textwidth]{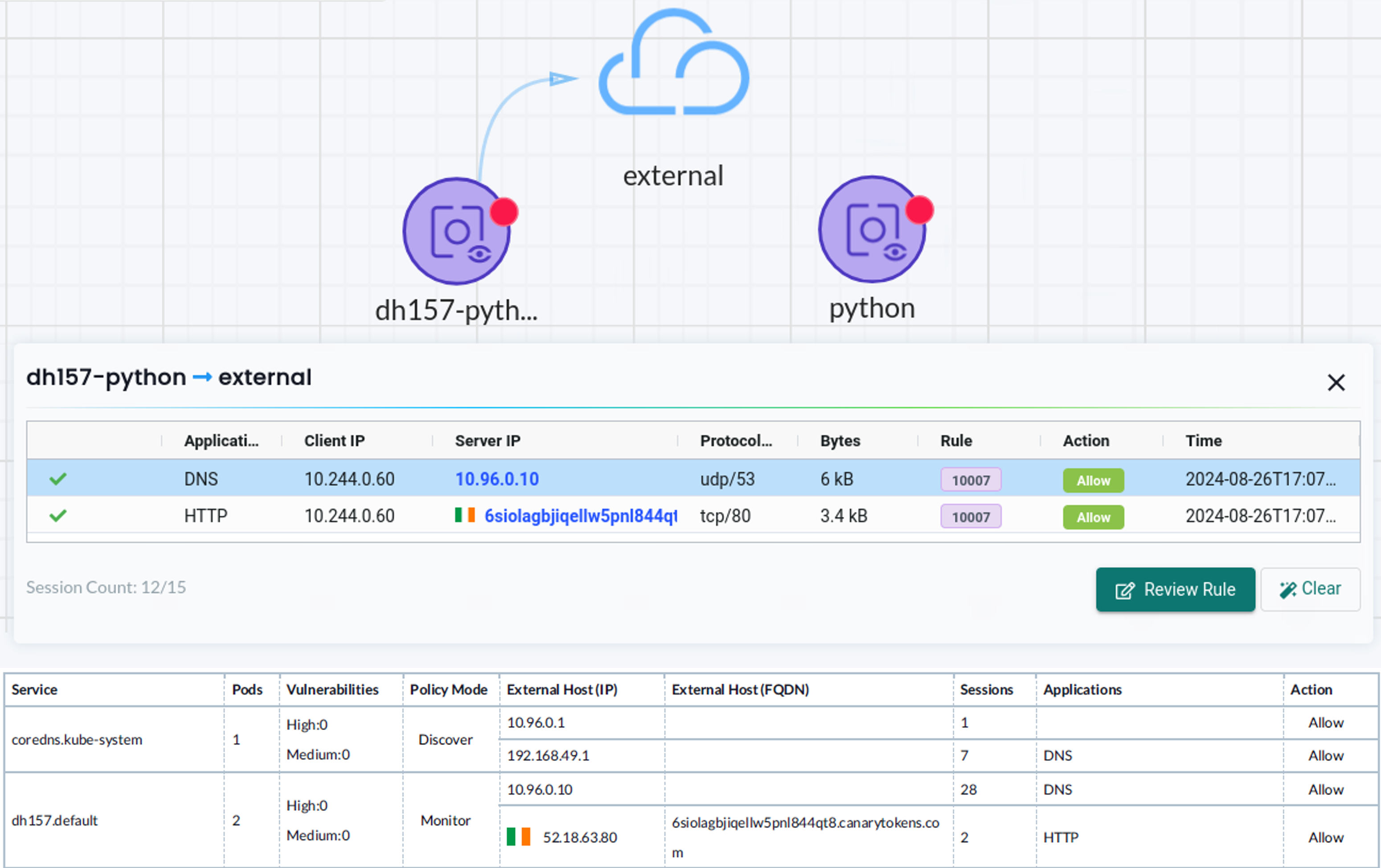}
    \caption{NeuVector Traffic Activity dashboard (top) and an extract from the summary report (bottom).}
    \label{fig:neuvector_discover_traffic}
\end{figure}

NeuVector is an open-source ``Full Lifecycle Container Security Platform''~\cite{gitneuvector}. Unlike the previous static scanning tools, NeuVector is a dynamic container security tool that provides features such as live traffic analysis, vulnerability analysis and scanning of deployed containers, and run time protection for both the containers and host system. NeuVector was deployed using Helm charts~\cite{helmcharts} via Minikube~\cite{minikube}. Once NeuVector was operational both the poisoned and original images were run within the NeuVector deployment. By default, NeuVector operates in the ``Discover'' mode, which learns the baseline behaviours of containers. The output of this mode raised no alerts for the poisoned image. Network traffic was observed and logged, but no security alerts or notifications were generated, as shown in Figure~\ref{fig:neuvector_discover_traffic}. 

Next, the deployment was restarted and set to ``Monitor'' mode before the the original or poisoned images were loaded and run. In this mode, NeuVector detected network traffic and raised a security alert, as by default the ``Monitor'' mode will alert on network activity to or from containers. However, the binary itself was not flagged within the container, with both the original and poisoned image reporting the same number of vulnerabilities. The only indicator in difference of behaviour of the two images was the network traffic. While it is not uncommon for containers and pods to transmit usage statistics, such as those collected by the popular Container Network Interface tools like Calico~\footnote{\url{https://docs.tigera.io/calico/latest/reference/resources/felixconfig}}, we wanted to conduct further testing to determine whether it would be possible to conceal this network traffic from detection while running the container in ``Monitor'' mode from the outset. 

To facilitate this additional testing, we created a new poisoned image with additional modifications to the binary (dh157-pid). In this test case, the binary sent a DNS request via the underlying host container. To achieve this, the container was run with privileged access and the host PID set to True (as shown in Code Listing~\ref{lst:dh157_pid}). 

\newpage


\begin{lstlisting}[caption={Run time parameters required for network evasion within NeuVector}, label={lst:dh157_pid}]
{
  "apiVersion": "v1",
  "spec": {
    "hostPID": true,
    "containers": [
      {
        "name": "dh157-pid",
        "image": "dh157/privileged:3.12-slim",
        "stdin": true,
        "tty": true,
        "command": ["/bin/bash"],
        "securityContext": {
          "privileged": true
        }
      }
    ]
  }
}'
\end{lstlisting}

\begin{figure}[!t]
\centering
\includegraphics[width=\textwidth]{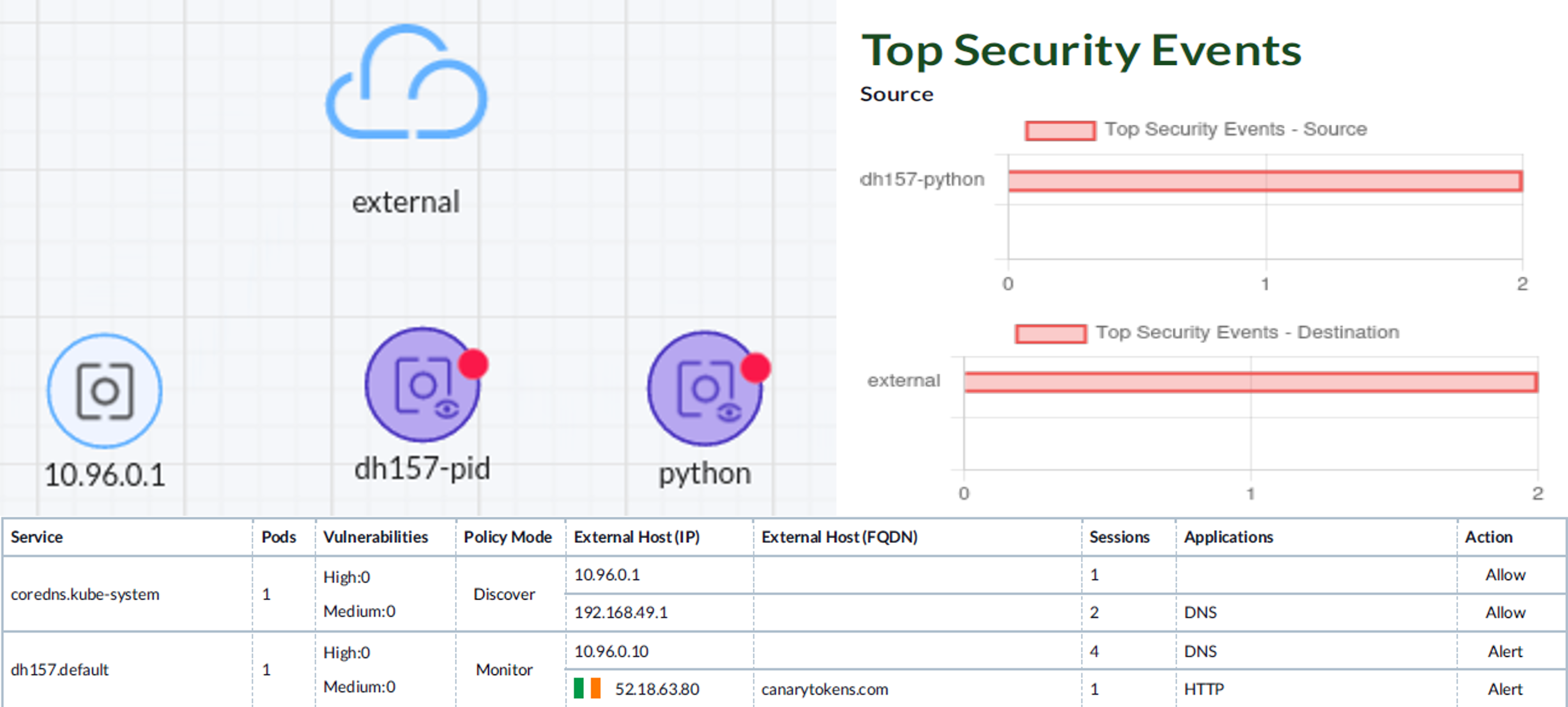}
    \caption{NeuVector Traffic Activity dashboard and Security Alert details (top) and an extract from the summary report (bottom).}
    \label{fig:neuvector_monitor_traffic}
\end{figure}

In this modified image and binary, a background task was launched using the {\tt nsenter}~\footnote{https://man7.org/linux/man-pages/man1/nsenter.1.html} command that executed a simple {\tt nslookup} bash command to generate traffic that would trigger a Canarytoken from within the host namespace whenever Python was called. This method successfully evaded detection by NeuVector, as the network traffic generated by the original poisoned image (dh157-python) was detected, while the traffic from the newly created image (dh157-pid) was not (Figure~\ref{fig:neuvector_monitor_traffic}). This concealment allowed our potentially malicious traffic to be transmitted undetected by NeuVector, and the expected alert from the Canarytoken was triggered.

\begin{figure*}[!t]
\centering
\includegraphics[width=0.9\textwidth]{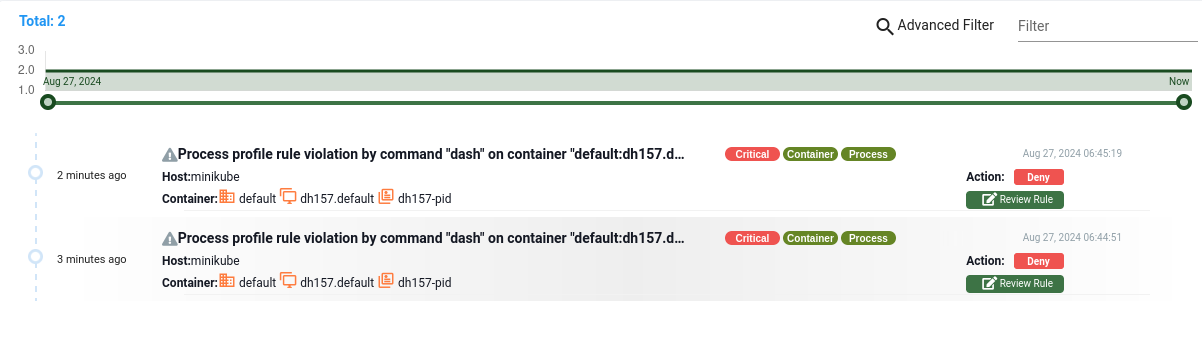}
    \caption{Alerts raised within NeuVector by the dh157-pid container}
    \label{fig:neuvector_protect_alerts}
\end{figure*}

The deployment was then terminated and restarted, this time only executing the newly created container dh157-pid. The deployment was configured to ``Protect'' mode before launching the container. Initially, the payload triggered a ``profile violation'' alert (as shown in Figure~\ref{fig:neuvector_protect_alerts}) due to a process decoupling, which allowed the {\tt nslookup} command to run independently from the container process. By removing the decoupling mechanism and threading, and executing the{\tt nslookup} command as a subprocess only, it became possible to run the container in a ``Protected'' mode deployment without raising any security alerts. 

\begin{figure*}[!t]
\centering
\includegraphics[width=0.9\textwidth]{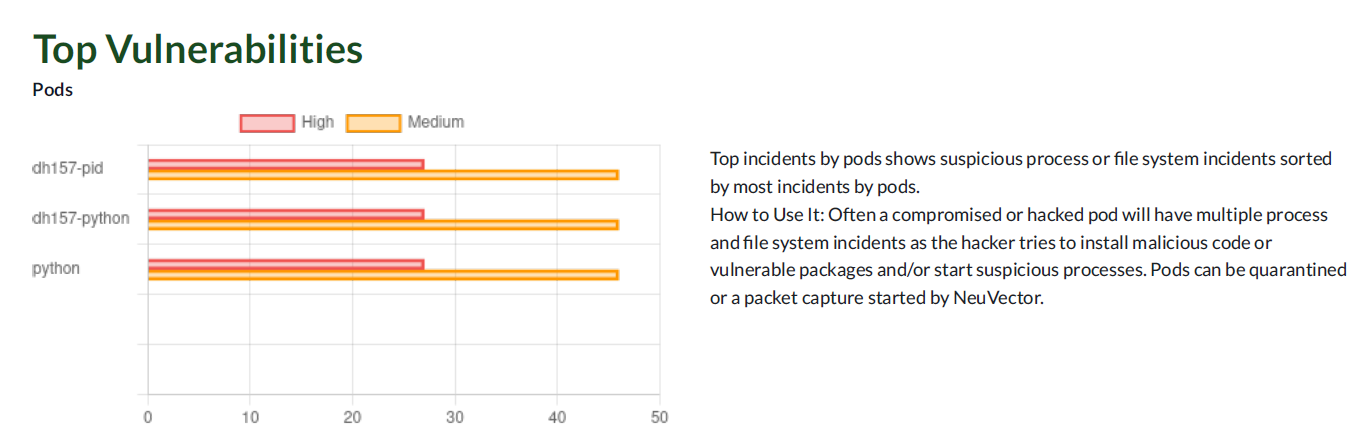}
    \caption{NeuVector Vulnerability reporting for the original and edited images}
    \label{fig:neuvector_vulns}
\end{figure*}

It is notable that all containers reported the same number of vulnerabilities (Figure~\ref{fig:neuvector_vulns}). This meant that once network traffic had been evaded there was no other indicators that would suggest any differences between the original image and the compromised image.

While this method requires an insecure container configuration, such settings could easily be hidden in a pre-provided deployment. This configuration is also not without legitimate use cases, as many applications require access to elements from the host system to work, such as the use of the Docker socket for tools like Dive. Therefore, it is feasible that a similar security-based application could be configured both with a privileged security context and and with access to the hostPID enabled. For instance, a containerised Intrusion Detection System (IDS) or process monitoring tool such as cAdvisor (Container Advisor) ~\footnote{\url{https://github.com/google/cadvisor/issues/2251}} would have a ``legitimate'' reason to be run under such circumstances. If such images were altered using gh0stEdit, they would be poisoned in a way that currently evades both static and dynamic analysis tools.

\subsection{Docker Trust}

To test our attack method against all recommended safeguards, we created a trusted and signed image based on an Alpine base image. This image was stored in a separate repository from all other poisoned images and was signed following Docker's guidelines for Docker Content Trust (DCT)~\cite{dockertrust}. The purpose of Docker Content Trust is to ensure the integrity of images, by signing an image manifest. The intent of DCT is to provide a level of trust for signed images, as stated by~\cite{dockertrust} ``[DCT] provides the ability to use digital signatures for data sent to and received from remote Docker registries. These signatures allow client-side or runtime verification of the integrity and publisher of specific image tags.''.

To determine whether the use of DCT would hinder our attack. We inspected the signature of the image before and after altering it by replacing the existing BusyBox with an older, vulnerable version of BusyBox, to also showcase a different attack vector. We successfully modified the image without any obvious impact or invalidation of the local signature. This is because there is currently no validation of the signed images digest against the local images digest. Only by conducting a full inspection of the image and comparing the \textit{RepoDisgest} value with that shown in the trust output would it be possible to identify that the image had been tampered with. This process would effectively require users to ``double check" the DCT output manually.  

We recorded this attack in its entirety and made the video available via our GitHub repository \url{https://github.com/amills157/gh0stEdit}, along with the full scan results from the attack examples detailed above.

\section{ Automated and Dynamic Attacks} 
\label{sec:automated_attack}

Building on this attack details covered in the previous section we created a simple bash script PoC that can automate the attack chain. This script can be used to pull down the latest version of an image, save it as a tar and extract it. Once extracted the script will find the latest layer to make a change to a specified directory (for example {\tt /usr/bin/local}), this then becomes the target layer for the attack and modification. Once the malicious edit has taken place the required manifest file is updated and a new .tar file created. This .tar is then loaded under the same name as the original image. As part of our automated experimentation we created a simple Rust reverse shell binary that was agnostic to the different Linux OS baselines (Alpine, RHEL, Debian, etc). The binary was also designed to take command line arguments which, after launching the reverse shell connection, it would then execute. To allow the automation of this attack, the name of the reverse shell binary ({\tt ghostedit\_rev\_shell}) was pre-pended to the image {\tt Command} or {\tt Entrypoint} as part of the gh0stEdit, so that any image downloaded would automatically execute the created Rust binary first, before continuing with the expected image execution. This allowed the gh0stEdit attack to be tested as part of an automated, dynamic attack chain in a easy and repeatable manner.

This attack script was tested against the images listed in Table~\ref{tab:base_images},as they had been identified as common base images across multiple sources (\cite{snykTop102019} and \cite{aimTop152024}) which represented different use cases and Linux OS.

\begin{table}[!t]
    \scriptsize
    \centering
    \begin{tabular}{ | c | } 
    \hline
       \textbf{Base Images} \\ \hline
        HTTPD \\ \hline
        Nginx \\ \hline
        Node \\ \hline
        Postgres \\ \hline
        Red Hat Universal Base Image \\ \hline
        Redis \\ \hline
        Ubuntu \\ \hline
    \end{tabular}
    \caption{Base images tested}
	\label{tab:base_images}
\end{table}

 A base image is something that is used to build other applications and services on top of. For example, using the Python base image to deploy bespoke, Python based, microservices. Therefore if the attack can successfully be carried out against these images, any image built on top of them will be equally vulnerable. Making such images viable targets, which themselves represent a very wide attack vector, for example, the Nginx image had been pulled over 9,000,000 times just between March 3rd and March 9th (2025).  Alpine and Python were also identified as popular base images, but has they had been used as part of attack cases 1 and 2, so were omitted from the automated attack.

\begin{figure*}[!t]
\centering
\includegraphics[width=0.9
\textwidth]{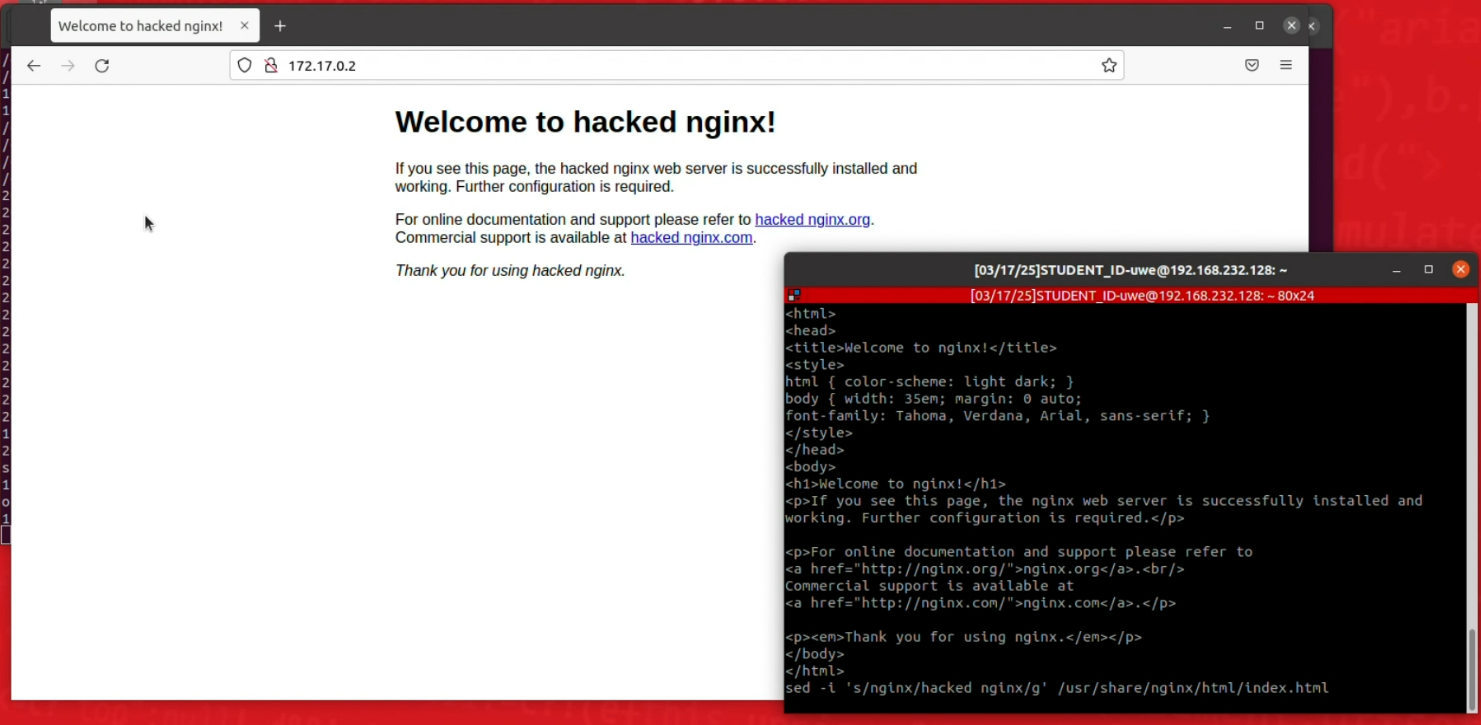}
    \caption{Malicious edit of the running Nginx container, via a gh0stEdit deployed RCE}
    \label{fig:gh0stEdit_Dynamic_Edit}
\end{figure*}

The attack was successful for all of the listed images, with the reverse shell connecting to the waiting listener, before continuing the normal execution of the image. This would give the attack access to the container files-system as the default user, often root. In the case of the Nginx image this allowed the attacker to dynamically alter the default index page whilst the container was running (see Figure~\ref{fig:gh0stEdit_Dynamic_Edit}).

This helps highlight how the gh0stEdit vulnerability can be exploited as part of a automated attack chain, for example in a CI/CD pipeline, which is discussed and expanded on further in ~\ref{sec:discussion}.

\section{Discussion} 
\label{sec:discussion}



The results from our attack examples and subsequent analysis demonstrate that this vulnerability can be exploited to compromise Docker images in ways that bypass both static and dynamic analysis. The only detectable changes in the edited images are the altered layer, image digest, and the increased image size. The image creation time, commands, and history remain unchanged, and newly added binaries or CVEs are not detected during scans. We have also demonstrated that by using gh0stEdit exploit, it is possible to create an image capable of sending out network traffic without raising any security alerts. The full scanner results can be found on our GitHub page\footnote{\url{https://github.com/amills157/gh0stEdit}}. 

The detectable indications of change rely on having a known `original' comparison image for reference. In a real-world attack scenario, such as a poisoned image in a Docker repository or an image maliciously altered within a development environment, detecting these changes would be extremely difficult. Image update are common, as seen in the official Python repository, which updated multiple images between 4th-8th and 13th-15th August 2024. These updates result in different image digests and layer IDs, particularly for images which are not tagged with a specific patch version, such as 3.12-slim or latest. In this scenario, once an image compromised using the gh0stEdit vulnerability is present in a repository, direct comparison becomes infeasible. The most reliable method of detecting such an attack would involve manual, in-depth layer-based analysis combined with reverse engineering and version checking of all installed binaries against the known SBOM.  

Within a production environment the exploitation of this vulnerability could occur at multiple stages. Manual exploitation could be carried out by a malicious insider with access to a repository at almost any stage. If a legitimate change is required, such as updating the code base for a container image, then gh0stEdit could be deployed simultaneously, to affect a malicious change that is masked by the legitimate update, providing ''cover" for pushing a new version of an image to a repository. 

The use of tags and labels \footnote{https://www.docker.com/blog/docker-best-practices-using-tags-and-labels-to-manage-docker-image-sprawl/} within production environments~\footnote{https://blog.nashtechglobal.com/docker-tagging-strategies-for-deploying-to-production} also provides another perfect ``cover" for the use of gh0stEdit. If image (re)tagging is automated within a CI/CD pipeline then a malicious edit to this pipeline will impact all images, potentially lead a supply chain attack on a massive scale. 

External attackers may also make use of gh0stEdit in combination with typo squatting or repository hijacking attacks. Making malicious edits to a legitimate image and then uploading it under similar but different names, confident in the knowledge that (currently) it will pass all industry recommended scanning practices, elevating this attack path to dangerous new heights compared to previous attempts which have been identified through the use of static scanning tools or image layer inspection~\cite{sysdig2022analysis}.

This type of attack evasion, combined with increased popularity of tools such as Kubernetes and Docker in DevOps~\cite{statistadevops2024}, provides a wide attack surface that requires a low level of technical complexity to achieve. The main mitigating factor for the successful use of gh0stEdit is access to either a trusted repository or deployment pipeline. However any developer operating on a process that open source images which pass static or dynamic scans, such as those covered in this paper, are safe to use, is at risk of a significant supply chain attack through the exploitation of gh0stEdit.

As gh0stEdit is itself a way to exploit an existing vulnerability within Docker images, the attacks that it can be utilised for are varied. As demonstrated as part of our case study it is possible to include malicious binaries which can connect out from the container image itself. This can take the form of CryptoMiners, a commonly deployed malware within container images, or binaries designed to exfiltrate logging or confidential data as part of a targeted attack (such as an insider threat). It could also be used to deploy backdoors or downloaders as part of a staged attack, however attackers would still be confined to the container environment, at least initially. Attacker motivations and targets may also vary, for example gh0stEdit could be used in combination with Kubernetes specific ransomware attacks~\cite{MITRESiloscape} or other destructive attacks. The main mitigation to such attacks is the potential for network traffic to be identified stopping a long running attack in its tracks. Something which can be overcome, but would require either insider knowledge of the targets defensive posture or a wide spread attack aiming to infect and capitalise on poor cyber security practice and defences.

The attack examples and analysis presented in this paper highlight significant vulnerabilities within the current container ecosystem and approaches to Docker-based security. Fundamentally, the recommended safeguards for container-based security rely on an assumed transparency and inherent trust in the container image. End users and scanning tools trust that the image hierarchy and self-reported SBOMs are accurate, a trust that gh0stEdit exploits. 

We have also demonstrated that the gh0stEdit exploit is effective on images signed as part of the Docker Content Trust (DCT) process, underscoring the severity of this attack and revealing issues within the DCT image signing process. To mitigate this risk, image signatures should be verified and invalidated locally to ensure signed images are not subsequently tampered with. Currently a signed image could be maliciously altered without any obvious indications of modification. In a CI/CD environment, this could result in the poisoned image being deployed as part of a production-ready asset, having passed all recommended safeguards.

\begin{figure*}[!t]
\centering
\includegraphics[width=0.9
\textwidth]{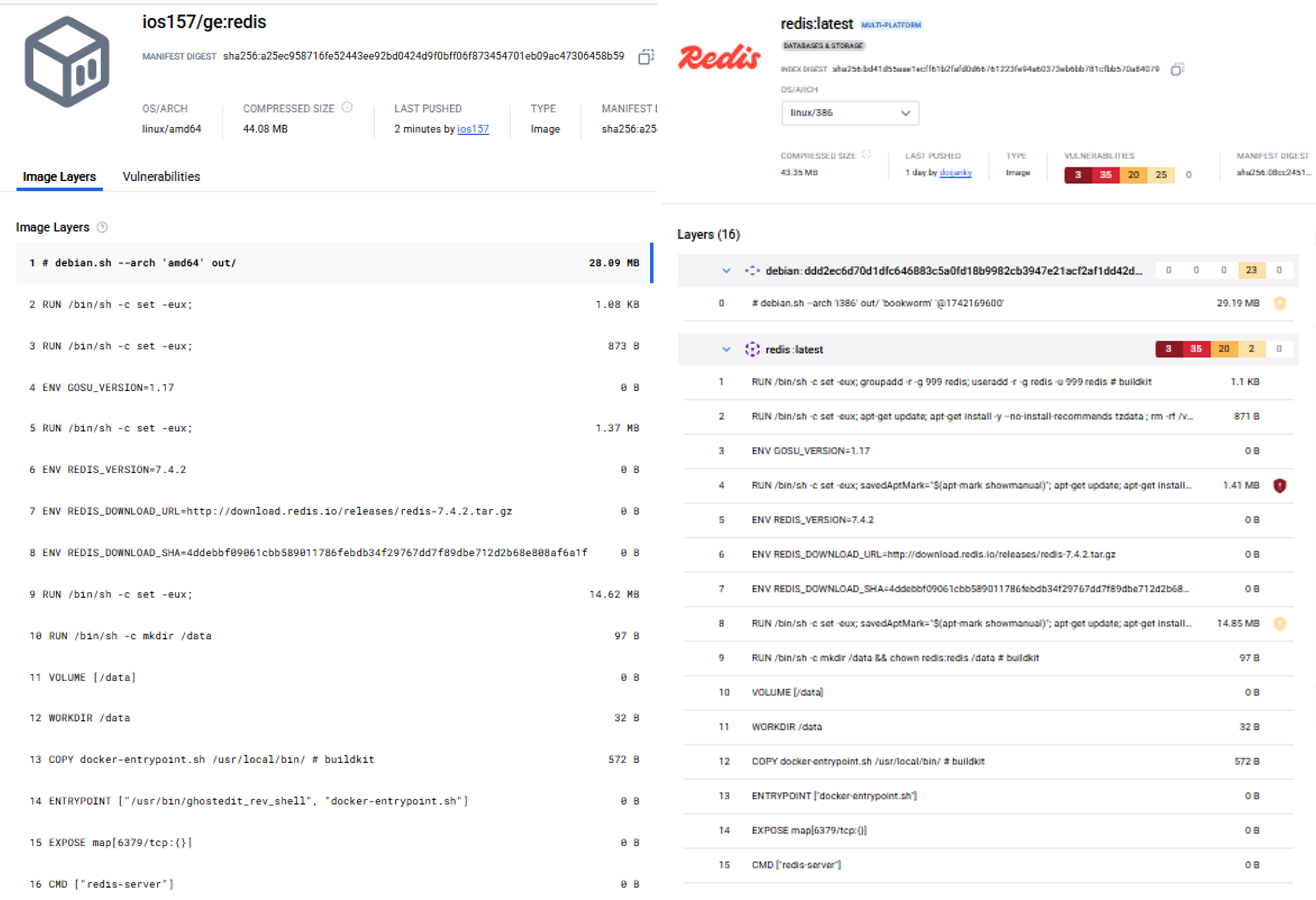}
    \caption{Comparison of edited and official Redis images - XFS gh0stEdit}
    \label{fig:Redis_XFS_Comparision}
\end{figure*}

\subsection{Cross Compatibility}

The primary environment used during the testing and investigation of gh0stEdit was an Ubuntu 20.04 LTS, which uses (by default) the EXT4 filesystem. Testing was however carried out within a CentOS 9 stream Virtual Machine, using the XFS filesystem. Within the XFS filesystem the structure of the docker images when saved and extracted differed, with additional nesting layers and JSON files. Whilst this meant our automated script did not work, having been designed for the EXT4 environment, manual testing of the gh0stEdit vulnerability was successful. Figure~\ref{fig:Redis_XFS_Comparision} shows the comparison between the Redis image which has been altered, using gh0stEdit, on an XFS system and the official image. The edited image had the reverse shell added, as described in~\ref{sec:automated_attack} and we can see that there are no obvious changes to either the layers or the image {\tt Command}.

\section{Responsible Disclosure and Ethical Considerations}

\label{sec:responsible_disclosure}

The gh0stEdit exploit was responsibly disclosed to Docker on the 14th of August 2024. Our disclosure included a Proof-of-Concept (PoC) bash script for editing the final layer within a Docker image, a video demonstration of the exploit, and an edited Docker image for analysis. All artefacts were shared privately with Docker to avoid third-party exposure or reverse engineering. Subsequent updates and further attack examples, such as the attack on signed images, were also communicated privately with Docker under their vulnerability disclosure policy~\footnote{\url{https://www.docker.com/trust/vulnerability-disclosure-policy/}}.

We acknowledge that the attack methods discussed in this paper could be misused, but we believe that publicly disclosure is necessary to raise awareness of these vulnerabilities. At present, no comprehensive measures are in place to prevent such attacks occurring. Without open discussion and demonstration, developing a solution will be unlikely. We encourage the academic and cyber security communities to act on these findings and work towards enhancing the security of the container ecosystem. 

\section{Conclusions and Further Work}
\label{sec:limitations}

In this paper we present gh0stEdit, the exploitation of a vulnerability in Docker images that enables attackers to compromise the integrity of an image. By saving an image as an archive file, an attacker can directly edit the layers, overwriting existing layers with malicious edits. Our case studies demonstrate that these edits are extremely difficult to detect and can bypass recommended industry-standard safeguards, including static and dynamic scanning, as well as image signing via Docker Content Trust (DCT). Using this exploit we successfully introduced a poison binary and intentionally added CVEs into images, which went undetected by multiple vulnerability scanners that are commonly used. We also showed that an image signed through DCT can be maliciously altered without invalidating its signature. To the best of our knowledge we are the first to publicly detail and discuss the exploitation of this vulnerability. 

This exploit presents a serious threat, particularly in the context of software supply chain attacks and invalidates existing trust in image transparency. These malicious edits do not appear in an image's history, hierarchy, commands, or SBOM. It would therefore be challenging to determine if this vulnerability has been exploited in practice without conducting an in-depth manual inspection of the image.

To mitigate against gh0stEdit, container image scanners and safeguards need to be restructured. Security checks that rely on self-reported data should be replaced with a ``zero trust'' approach. Content must be directly examined to ensure installed packages are free of known vulnerabilities, rather than relying on self-reported SBOMs.

One possible solution would be the introduction of an ``image or layer edit time'' field, which would be reported alongside existing image metadata, such as the image creation date. This would ensure that any post-creation changes to the container are evident, and could mitigate against gh0stEdit.

Additionally, the logic behind image signing and Docker Content Trust needs to include a verification check between the signed \textit{RepoDigest} and the digest of the inspected image. This would ensure that any changes made to an image would invalidate its signed status, providing an additional layer of security.


\section*{Acknowledgements}

This work was supported by the College of Arts, Technology and Environment at the University of the West of England.

\bibliographystyle{elsarticle-num} 
\bibliography{references}

\end{document}